\documentclass[prb,onecolumn,nobibnotes,groupedaddress, natbib]{revtex4}

\usepackage{enumerate}
\usepackage{amssymb}
\usepackage{amsmath}
\usepackage{mathtools}
\usepackage{amssymb}
\usepackage{braket}
\usepackage[hidelinks]{hyperref}
\usepackage{pdfpages}
\usepackage{epstopdf}

\begin{document}

\title{An assessment of frozen natural orbitals and band gaps using equation of motion coupled cluster theory: a case study on polyacene and trans-polyacetylene}

\author{Zachary W. Windom$^{1, 2}$\footnote{\href{mailto:zww4855@gmail.com}{zww4855@gmail.com}}, AV Lam$^1$, Ajith Perera, and Rodney J. Bartlett$^1$}
\affiliation{$^1$Quantum Theory Project, University of Florida, Gainesville, FL, 32611, USA \\
$^2$Computational Sciences and Engineering Division,\ Oak\ Ridge\ National\ Laboratory,\ Oak\ Ridge,\ TN,\ 37831,\ USA}
%\ead{support@elsevier.com}

\begin{abstract}
Frozen natural orbitals (FNOs) are used to augment IP/EA-EOM-CCSD calculations targeting the band gap of trans-polyacetylene and polyacene.
We show the resulting electron affinities (EAs), ionization potentials (IPs), and extrapolated band gaps incur errors that are largely tunable  to a desired accuracy,  yet require many orders of magnitude fewer core-hours as compared to the corresponding full calculation.  The relationship between various FNO truncation schemes and (cc-pV$n$Z) basis set is also examined. %Instabilities present in the reference wavefunction and the existence of adjacent IP/EA excited surfaces are found in large oligomers of polyacene, which we identify as the culprit behind the band gap discontinuity reported in prior work. %using mean-field and single-reference methods.

%A hybrid method to extract the band gap is further proposed that modifies large basis EOM-MBPT(2) results by incorporating missing correlation recovered in smaller basis set EOM-CCSD results.

%The accurate, $ab initio$ prediction of band gaps for one-dimensional, $\pi$-conjugated systems continue to be an area of active research. In this work, we employ ionization potential (IP) and electron affinity (EA) equation of motion coupled cluster calculations at the singles and doubles level (IP/EA-EOM-CCSD) in conjunction with frozen natural orbitals (FNOs) to target the fundamental band gap for unit cell geometries of trans-polyacetylene and polyacene. We extrapolate to the infinite, polymer-limit band gap by studying oligomers of increasing size. Comparison against the full IP/EA-EOM-CCSD results show that -------
\end{abstract}

\maketitle

\section{Introduction}

%Why band gap is important
The fundamental band gap is formally defined as the energy required to promote an electron from the highest occupied to the lowest unoccupied band,\cite{ashcroft1993solid} and further governs the electronic properties of extended systems.\cite{kittel2021introduction} Consequently, it is a quantity of fundamental interest. Efforts focused on characterizing the band gap of $\pi$-conjugated polymers have soared recently. This is at least in part attributable to their potential use in optoelectronic\cite{ostroverkhova2016organic,wang2014flexible} applications as well as in many biological contexts.\cite{liu2023optical} The prototypical examples of such systems are polyacene and trans-polyacetylene, which are also of current interest in this work. In order to accurately survey these polymers \emph{ab initio} and \emph{in silico}, the employed electronic structure method must be tractable and have a high degree of fidelity . 

One possible route in extracting the band gap is based on a periodic boundary conditions approach using either planewaves or Gaussian orbitals to directly probe infinite limit behavior.\cite{sun2018pyscf} Several modern software packages\cite{valiev2010nwchem,giannozzi2009quantum} offer these capabilities, particularly in the context of Kohn-Sham density functional theory (KS-DFT).\cite{hohenberg1964inhomogeneous,kohn1965self} An alternative, ``bottom-up" approach would employ more reliable wavefunction methods to observe trends in oligomers of increasing size, then extrapolate to the infinite limit. Based on its documented history of success guaranteeing size-extensive results for the ground state,\cite{hirata2001highly} coupled cluster theory (CC)\cite{bartlett1981many} and its equation of motion extension for excited states (EOM-CC)\cite{stanton1993equation} that ensure size-intensivity for energy differences seem ideally suited for this task. Unfortunately, calculations employing the standard CC ansatze are computationally expensive; although various approximations have been proposed that show promise in overcoming these challenges.\cite{flocke2003correlation,flocke2004natural,hughes2008natural,jin2018perturbation,riplinger2013efficient,mcclain2017gaussian}

% Describe FNO-based approach - prior work --- literature review 
A potential way to circumnavigate these expenses is to utilize a set of orbitals designed to exhibit superior convergence characteristics. The Natural Orbitals fit such a criteria, where it has been shown that a fraction of  canonical, configuration state functions are required to ensure rapid convergence to the FCI limit.\cite{lowdin1956natural} This concept has been adapted for methods based on CC theory, leading to the so-called Frozen Natural Orbitals (FNOs), wherein a set of occupation numbers are retrieved by constructing and diagonalizing the virtual orbital sector of the MBPT(2) density matrix, keeping the occupied orbitals fixed.  Eigenvectors corresponding to ``large" virtual orbital occupation numbers, as governed by some user-defined criteria, then define a set of FNOs.\cite{taube2005frozen}  The eigenvectors which correspond to ``negligible" occupation numbers are deemed less important and typically disregarded in subsequent calculations. Studies have shown that this basis set truncation scheme can be applied without significant impact to CC total energies.\cite{taube2008frozen} As a byproduct of reducing the basis set rank, subsequent CC calculations can be accelerated and significantly larger molecular complexes can be studied.\cite{byrd2014chain} Clearly, the synergy in using FNOs alongside ground-state CC theory is a potential route toward making CC-quality results routinely accessible to large-molecule studies.  

However, the current work uses FNOs in a different context; namely to construct the target states in EOM excited states. Any EOM-CC calculation has two components: (1) the ground or reference state solution defined by exp(T) which is encapsulated in the effective CC Hamiltonian, $\overline{H}$, and (2) the target state solution defined by $\overline{H}R_k=\omega_kR_k$. As the ground state is common, taking an MBPT2 approximation to generate FNO's for EOM-CC is meaningful, but it is not specific to a particular target state. For IP's the similarity between the ground state's orbital structure and its principal ionizations make this problem relatively simple, but that is not necessarily true for most other EOM targets, like EAs, EEs, DIPs, DEAs, etc.
To this end, the FNO framework has already proven capable of targeting both ionized states,\cite{landau2010frozen} open shell systems,\cite{pokhilko2020extension} and linear response properties.\cite{kumar2017frozen} Because the reference space of the EOM $\overline{H}$ matrix includes 1h terms which are the largest contributor to ionization, FNO-based IP-EOM-CCSD is expected (and already been shown) to yield results that can be tuned to a desired accuracy according to the truncation level. 

On the other hand, the usual target state description based on MBPT2 ground state FNOs is more ambiguous for both EAs and EEs. Regarding EAs, this is because the ground MBPT2 reference space does not include the largest EA contributor, 1p attached states. Nevertheless, we show that using a MBPT2 approximation to capture a single-reference CC EA state can be largely tailored to the problem at hand. Limited error associated with this incompatibility is expected for FNO-EAs as compared to FNO-IPs, although a full accounting of the relative errors differentiating the two are largely absent in the literature. Here, we intend to address this by  benchmarking both FNO-based EOM calculations.

To this end, we augment prior efforts\cite{windom2022examining} by incorporating FNOs into band gap calculations of polyacene and trans-polyacetylene, using ionization potential (IP) and electron affinity (EA) EOM-CC calculations at the singles and doubles excitation level (IP/EA-EOM-CCSD). Errors associated with truncating the virtual space at various thresholds are reported. %We further show the efficacy in employing IP/EA-EOM-MBPT(2) (many body perturbation theory second order)  calculations. Since EOM-MBPT(2)  significantly reduces the dimensionality of the matrix requiring iterative diagonalization, comparatively larger basis sets are accessible to EOM-MBPT(2) that would otherwise remain prohibitively expensive. 

The utility of FNO-based EOM calculations is further emphasized in a study of the band gap discontinuity of polyacene reported in prior work\cite{windom2022examining} using Kohn-Sham Density Functional Theory (KS-DFT). With the aid of FNOs, EOM calculations on polyacene oligomers up to the 14mer are possible. It was thus discovered that two excited surfaces were found to  exist in close proximity; a result that would have been exceedingly difficult to find without the aid of FNOs in conjunction with high-level wavefunction theory.

% prior work on ionized states & open shells. 

% What we do in this work: FNO calcs on protypical conjugated polymers. Why this info on these polymers is important

It should be heavily emphasized that every calculation in the current work utilizes a single CPU, with a maximum walltime of 396 core-hours for each job unless noted otherwise. This is in stark contrast to prior work\cite{windom2022examining} which required use of the massively parallel ACESIII\cite{deumens2011software} and up to 576 CPUs at a time, although the use of point group symmetry was admittedly absent in those calculations. Whereas these results were limited to roughly 800 basis functions, the current work exceeds 1300 basis functions in some instances and requires a fraction of the core-hours. %This fact alone perhaps argues for a reconsideration in how band gaps are extracted using wave function theory. 

%, thereby offering an third avenue for obtaining the thermodynamic-limit band gap. 

\section{Computational methodology}

%% Geometry/Calculations

 Average geometries were constructed for polyacene and trans-polyacetylene, based on the respective unit-cells; the geometry of the former is fully aromatic whereas the latter emulates experimental bond length and angle parameters.\cite{yannoni1983molecular} All calculations utilize the ACESII software\cite{perera2020advanced} and full point group symmetry, given to be $D_{2h}$ for polyacene and $C_{2h}$ for trans-polyacetylene. The FNO errors are determined with respect to the full IP/EA-EOM-CCSD calculations catalogged in prior work,\cite{windom2022examining} and retain 10, 20, 30, 40, 50, 55, 60, 65, 75, and 85\% of the virtual space. Virtual space truncation was based on the MBPT(2) natural orbital occupancy constructed from the first-order density matrix. All EOM-CC band gap calculations employ the cc-pVDZ and cc-pVTZ basis sets\cite{dunning1989gaussian} for polyacene and trans-polyacetylene, respectively. For analysis of the basis set dependence on the  FNO truncation scheme, extrapolated IPs utilize a formula to the complete basis set (CBS) limit using the equation $A(X)=A(\infty)+Be^{-(n-1)}+Ce^{-(n-1)^2}$ recommended by Peterson \textit{et. al.} in prior work.\cite{peterson1994benchmark,balabanov2006basis} All calculations drop the core orbitals and use the core Hamiltonian as the initial guess to the SCF. The SCF wavefunction is checked to verify RHF$\rightarrow$RHF instabilities are not present. %\textbf{(Zack. Isn't this contrary to the prior paper where we say instabilities occurred even for 3-mers. But that is a general UHF instability I assume not a RHF to RHF one? Clarify)}

Bandgap extrapolations follow standard convention by optimizing the function $Ae^{-\frac{n}{B}}+C$ for $n$ unit cells, and are performed on the interval 6-8 for polyacene and 6-9 for trans-polyacetylene. The absence of data for pentacene (all FNO calculations) and the nonacene (FNO 10, 20) reflect instances where the CC equations did not converge. Similarly, extrapolation data is absent for trans-polyacetylene at FNO 85 as some calculations did not converge within the allotted walltime (8,9mer). All plots and analysis utilize the numpy,\cite{harris2020array} matplotlib,\cite{hunter2007matplotlib} and SciPy\cite{virtanen2020scipy} Python libraries.

To further analyze the band gap discontinuity in polyacene, oligomers as large as the 14mer were studied using a FNO approximation that keeps 60\% of the virtual space. We modify the irreducible representation distribution of occupied orbitals at the SCF level to uncover two separate IP/EA surfaces that are in close proximity, particularly at larger oligomer lengths. These two distributions, denoted as Occ. Set 1 and 2, represent local minima on the potential energy surface as verified by stability analysis. For all polyacene calculations, we constrained the wavefunction to $D_{2h}$ symmetry where the ground state is $^1A_g$.

%In addition to using the aforementioned basis sets to quantify error, additional EOM-MBPT(2) calculations were performed that employ larger basis sets: cc-pVTZ for polyacene and cc-pVQZ for trans-polyacetylene. As a third way to quantify the band gap, we record a correction to the larger basis EOM-MBPT(2) results by augmenting the ground state correlation energy with the CCSD energy. Further details on the chosen geometries, extrapolation formula, and the ``hybrid" method are provided in the supplementary information. All plots and analysis use the numpy, matplotlib, and SciPy Python libraries. ***CITE****

\section{Results}
%Discuss error a little further
The FNO error for IPs and EAs of both polyacene and trans-polyacetylene are quantified in Tables 1 and 2, and visualized in Figures 1 and 2, respectively. As expected, truncating the virtual space by varying extents leads to IP errors that are largely systematic. In fact, truncating 60\% and 35\% of the virtual space of trans-polyacetylene and polyacene, respectively, reduces the IP error to 0.1 eV of the corresponding full calculation. On the other hand, obtaining an error of 0.1 eV for the EA requires as much as 85\% of the virtual space to be kept but is nevertheless obtainable in both systems. In spite of this, truncating between 35-40\% of the virtual space generates extrapolated band gap results that are within 0.1 eV of the full calculation.

 \begin{figure}[ht!]%{0.6\textwidth}
 \centering
\includegraphics[scale=0.15]{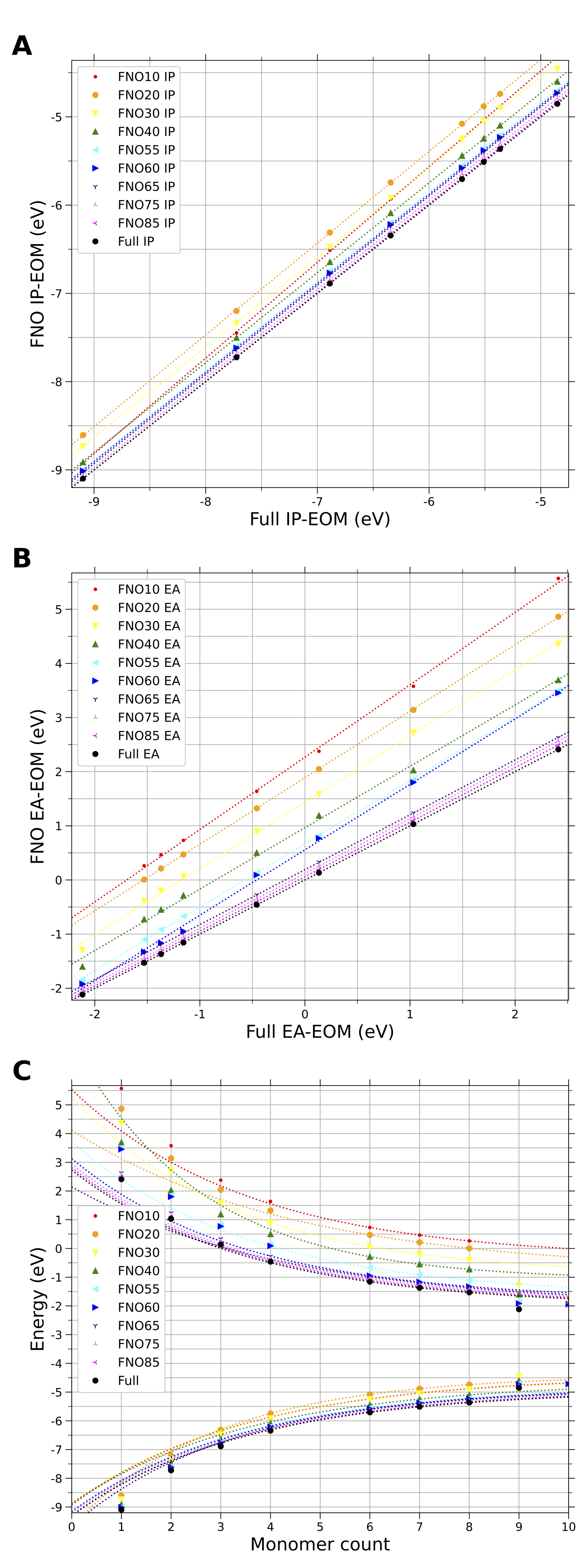}
\caption{A) Comparison of FNO-based IP-EOM-CCSD against full calculation, B) comparison of FNO-based EA-EOM-CCSD against full calculation  and C) FNO-based band gap predictions of polyacene.}
\label{fig:polyaceneIMG}
\end{figure}

\begin{figure}[ht!]%{0.6\textwidth}
\centering
\includegraphics[scale=0.15]{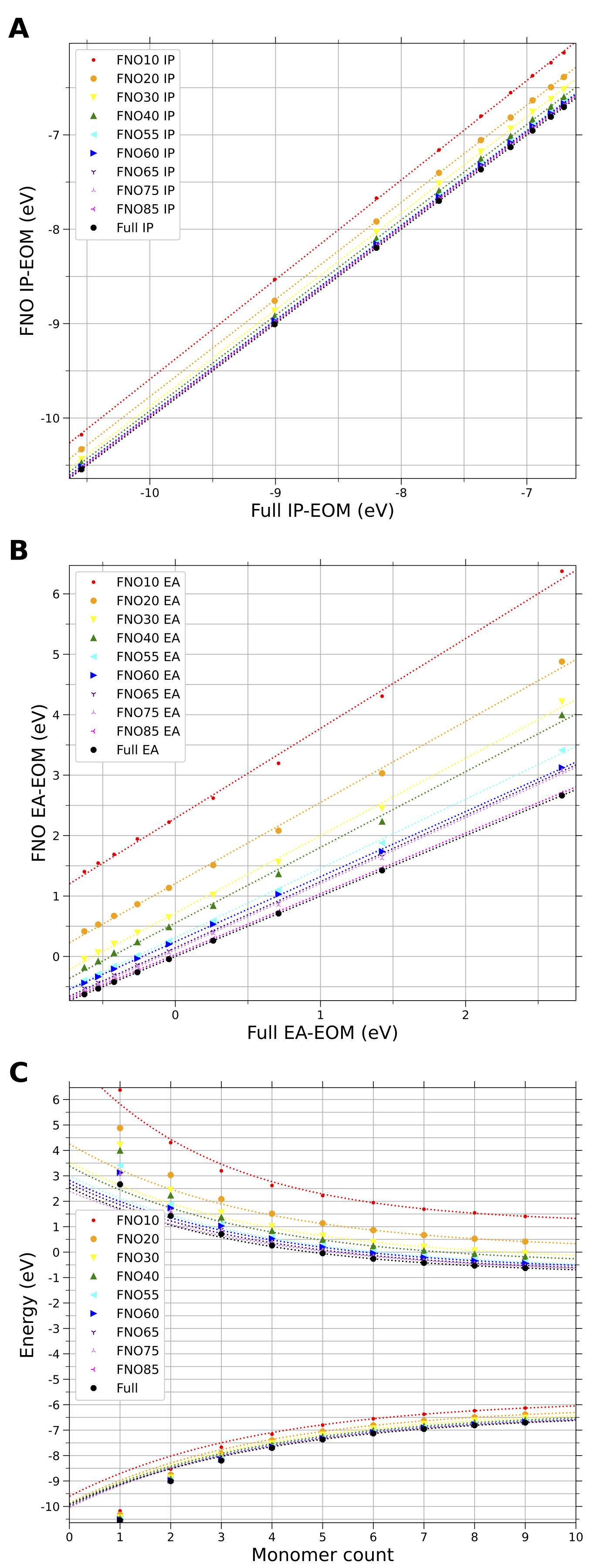}
\caption{A) Comparison of FNO-based IP-EOM-CCSD against full calculation, B) comparison of FNO-based EA-EOM-CCSD against full calculation  and C) FNO-based band gap predictions of trans-polyacetylene.}
\label{fig:tPAIMG}
\end{figure}

\begin{table}
\begin{center}
    \caption{Error comparison of various FNO approximations against the corresponding full EOM-CCSD calculation for trans-polyacetylene, specifically the signed error (SE) and mean signed error (MSE). Extrapolations absent for FNO85 are attributable to calculations of the 9mer exceeding the 396 core-hour walltime.}
    \label{tab:tPAFNOerror}
\begin{tabular}{  c c c  c c}\hline
	&	Extrapolated band gap	&	Extrapolated band gap SE	&	EA MSE	&	IP MSE \\ \hline
FNO10	&	6.91	&	1.49	&	2.46	&	0.53 \\
FNO20	&	6.11	&	0.68	&	1.33	&	0.29\\
FNO30	&	5.78	&	0.36	&	0.82	&	0.17\\
FNO40	&	5.74	&	0.31	&	0.64	&	0.10\\
FNO55	&	5.60	&	0.17	&	0.36	&	0.05\\
FNO60	&	5.56	&	0.14	&	0.27	&	0.04\\
FNO65	&	5.49	&	0.07	&	0.18	&	0.03\\
FNO75	&	5.45	&	0.03	&	0.14	&	0.02\\
FNO85	&	-	&	-	&	0.04	&	0.02\\
Full	&	5.42	&	-	&	-	&	-\\ \hline
\end{tabular}
\end{center}
\end{table}

\begin{table}
\begin{center}
    \caption{Error comparison of various FNO approximations against the corresponding full EOM-CCSD calculation for polyacene, specifically the signed error (SE) and mean signed error (MSE).}
    \label{tab:paFNOError}
\begin{tabular}{  c c c  c c}\hline
	&	Extrapolated band gap	&	Extrapolated band gap SE	&	EA MSE	&	IP MSE	\\ \hline
FNO10	&	4.07	&	1.16	&	2.22	&	0.38	\\
FNO20	&	3.58	&	0.67	&	1.86	&	0.58	\\
FNO30	&	3.52	&	0.62	&	1.43	&	0.43	\\
FNO40	&	3.53	&	0.63	&	0.97	&	0.24	\\
FNO55	&	3.08	&	0.18	&	0.65	&	0.14	\\
FNO60	&	2.97	&	0.07	&	0.52	&	0.12	\\
FNO65	&	2.83	&	-0.08	&	0.19	&	0.10	\\
FNO75	&	2.94	&	0.03	&	0.12	&	0.05	\\
FNO85	&	2.93	&	0.03	&	0.06	&	0.01	\\
Full	&	2.91	&	-	&	-	&	-	\\ \hline
\end{tabular}
\end{center}
\end{table}
% discuss discontinuity of PBE EOM --- cite this figure in SI
There are two things worth mentioning: 1) The closed-shell, reference wavefunction of polyacene transitions to singlet diradical character as the oligomer legnth increases, as indicated by the presence of broken spin-symmetry RHF$\rightarrow$UHF instabilities, verified in prior work.\cite{bendikov2004oligoacenes} 2) Upon further investigation of the band gap discontinuity which starts at nonacene, we found evidence that at least two local minima exist at the RHF level. Either minima can be accessed by appropriately constraining the population of occupied orbitals according to their irreducible representation. These two distributions, given in the supplementary material, are referred to as Occ. Set 1 and Occ. Set 2. 

\begin{figure}[ht!]
\includegraphics[scale=0.42]{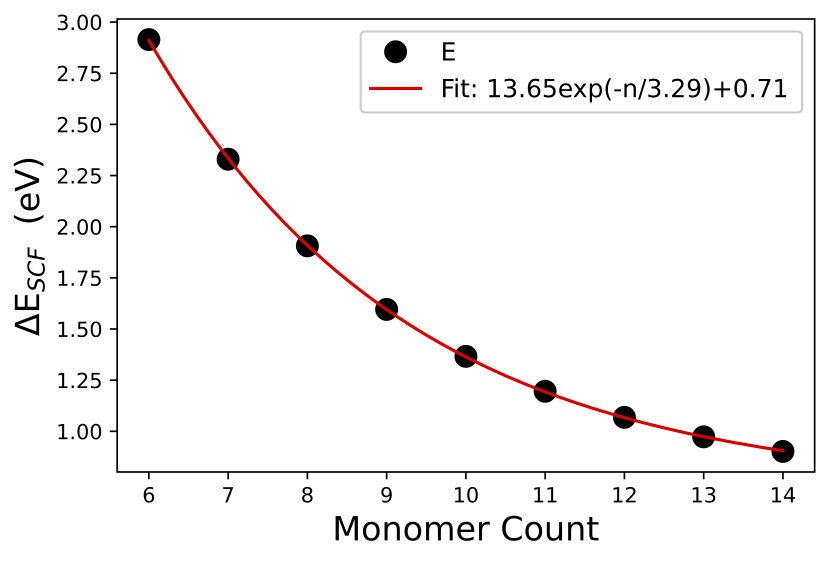}
\caption{Depiction of difference in SCF energy between Occ. Set 2 and Occ. Set, $\Delta E_{SCF}$, as a function of oligomer size. It is noted that the variational solution for Occ. Set 1 is predicted to be $\approx$ 0.71 eV below that of Occ. Set 2 in the infinite polymer limit. }
\label{fig:SCFenergy}
\end{figure}

Figure \ref{fig:SCFenergy} tracks the energy of each SCF solution up to the 14mer, and verifies that Occ. Set 1 is consistently lower in energy than Occ. Set 2. Extrapolating to the infinite polymer limit, this continues to be the case. The fit over the entire interval indicates the SCF energy of Occ. Set 1 to be 0.71 eV below Occ. Set 2, which is revised to 0.69 eV when the interval only includes the largest 4 oligomers. From this, we conclude that the choice in fitting interval does not significantly impact these results.

\begin{figure}[ht!]%{0.6\textwidth}
\includegraphics[scale=0.35]{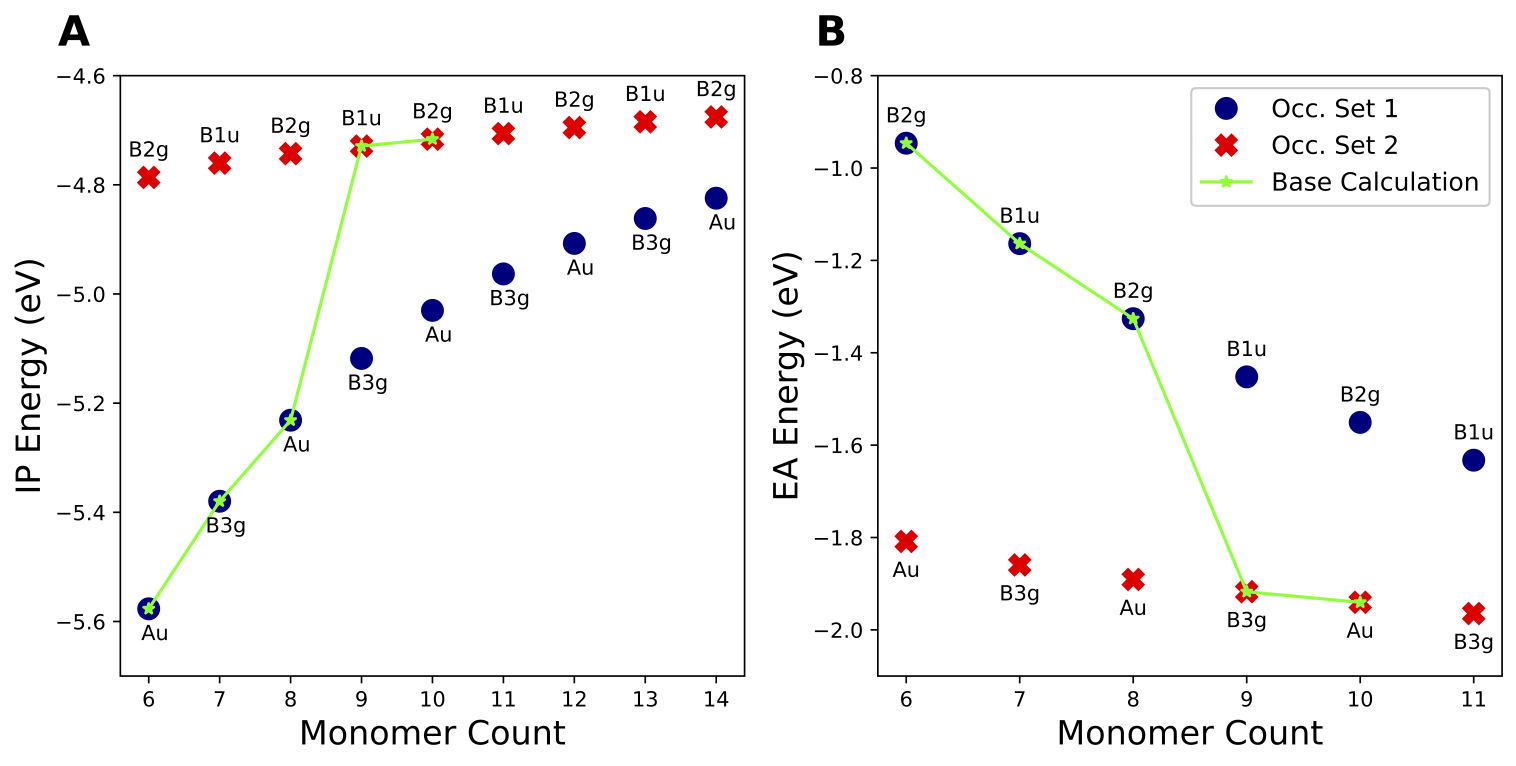}
\caption{Comparison of FNO-based (A) IP-EOM-CCSD and (B) EA-EOM-CCSD calculations of polyacene oligomers where the RHF SCF is constrained to have two different irrep distributions of the occupied orbitals. The green line ("Base Calculation") follows EOM-CCSD calculations wherein the RHF SCF is unconstrained. The irrep for the principal IP/EA is annotated.}
\label{fig:combinedIPEA}
\end{figure}

It is found that by appropriately constraining the initial population of occupied orbitals to Occ. Set 1, the discontinuity of polyacene reported in prior work can be resolved. In fact, it turns out that the band gap discontinuity is a byproduct of the unconstrained SCF transitioning from Occ. Set 1 to Occ. Set 2, which initially happens at nonacene.  This is further emphasized in Figure \ref{fig:combinedIPEA}, which compares the IP/EA surfaces that are generated by different choices in occupation set. We note that as the oligomer grows in size, these adjacent surfaces are in close proximity. 

. %The population distribution of the occupied orbitals according to irreducible representation defining Occ. Sets 1 and 2 are recorded in Tables 1 and 2 of the supplementary material, respectively.

\begin{figure}[ht!]%{0.6\textwidth}
\includegraphics[scale=0.32]{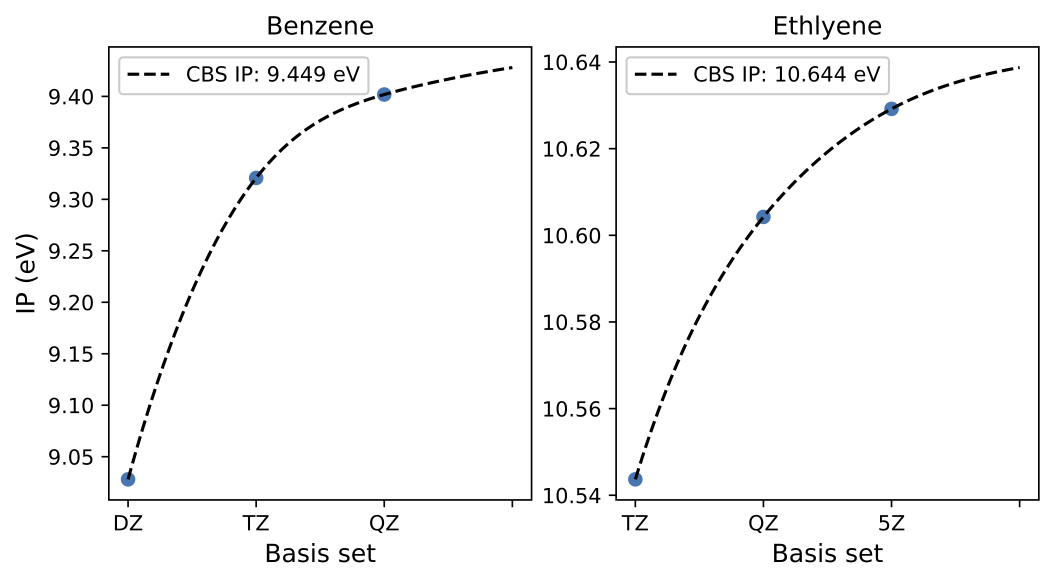}
\caption{Comparison of the CBS IP limit for benzene and ethlyene using IP-EOM-CCSD. }
\label{fig:fullCalcs}
\end{figure}

\begin{figure}[ht!]%{0.6\textwidth}
\includegraphics[scale=0.32]{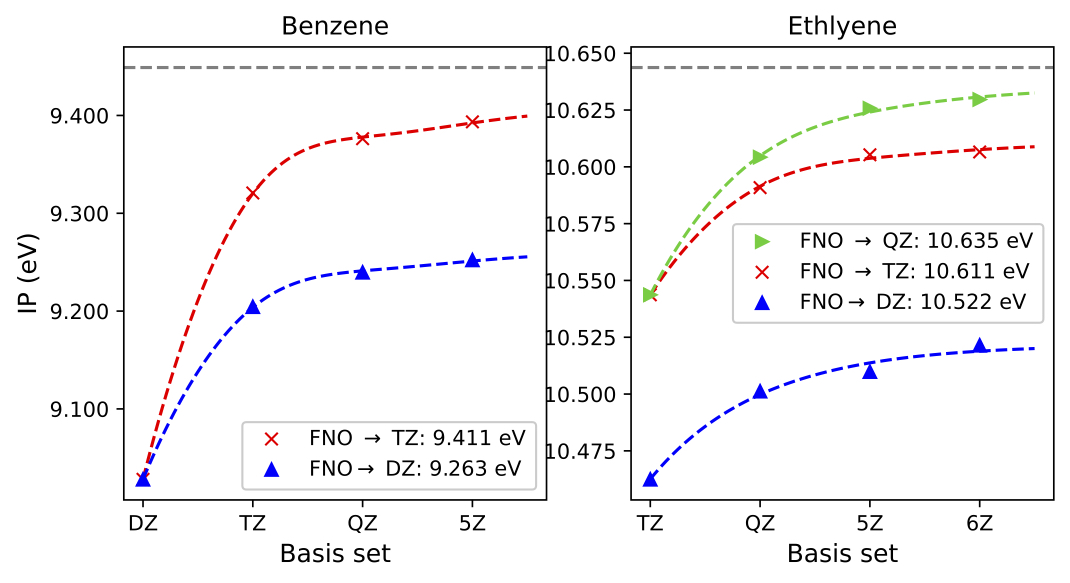}
\caption{Truncating the basis set dimension to DZ (FNO$\rightarrow$DZ), TZ (FNO$\rightarrow$TZ), or QZ (FNO$\rightarrow$QZ) using respective FNOs. The CBS limit estimated in Figure \ref{fig:fullCalcs} is shown by the grey dashed line.  Approximated FNO IP fit uses every point shown. }
\label{fig:truncateBasis}
\end{figure}

% discuss exhausting of the basis set
Another interesting question is accessible when using FNOs: how quickly do we exhaust the correlation effects within a particular basis set? To get a general idea of what the complete basis set (CBS) limit is without any approximation, Figure \ref{fig:fullCalcs} compares the systematic basis set convergence of the IP for benzene and ethlyene using the extrapolation formula of Peterson et al. Figure \ref{fig:truncateBasis} then compares the full value (shown by the grey dashed line) to the CBS limit IP which is generated using FNOs that truncate the total dimensionality of the basis sets used in the extrapolation procedure. For benzene, we analyze the CBS limit IP using two FNO options:
\begin{itemize}
    \item FNOs were generated keeping 38\%, 19\%, and 11\% of the virtual orbitals (VOs) to ensure the TZ, QZ, and 5Z basis sets, respectively, were of comparable size to the DZ basis set
    \item FNOs were generated keeping 50\% and 29\% of the VOs to ensure the QZ and 5Z basis sets, respectively, were of comparable size to the TZ basis set
\end{itemize} whereas for ethlyene we study three FNO options:
\begin{itemize}
    \item FNOs were generated keeping 37\%, 18\%, 10\%, and 7\% of the VOs to ensure the TZ, QZ, 5Z, and 6Z basis sets, respectively, were of comparable size to the DZ basis set
    \item FNOs were generated keeping 49\%, 28\%, and 17\% of the VOs to ensure the QZ, 5Z, and 6Z basis sets, respectively, were of comparable size to the TZ basis set
    \item FNOs were generated keeping 57\% and 35\% of the VOs to ensure the 5Z and 6Z basis sets, respectively, were of comparable size to the QZ basis set
\end{itemize}
It is clear that keeping more of the VOs in a FNO approximation results in better extrapolations that systematically converge toward the corresponding full CBS limit IP, as expected. In either example, ensuring the basis sets in the extrapolation procedure are no larger than TZ results in an error of 0.03 eV, which increases to $>$0.12 eV when the basis sets are constrained to have a DZ dimensionality. For ethlyene, 0.009 eV agreement can be found with the fully extrapolated answer if the basis set dimensionality is constrained to being no larger than QZ. 

To emphasize the saving in computational cost between the full and FNO-based approaches, it is noted that the full 6Z IP-EOM-CCSD calculation \textbf{exceeds 100 core-hours}. On the other hand, when using FNOs that reduce the dimensionality of the 6Z basis set down to the size of QZ, the corresponding IP calculation achieves chemical accuracy with the unapproximated, full IP \textbf{but requires only 5 core-hours}. Clearly, there is little downside to using FNOs to extrapolate toward the basis set limit in the case of benzene and ethlyene. However, it is possible that increasing oligomer size introduces more error into the FNO approximation. 

\begin{figure}[ht!]%{0.6\textwidth}
\includegraphics[scale=0.25]{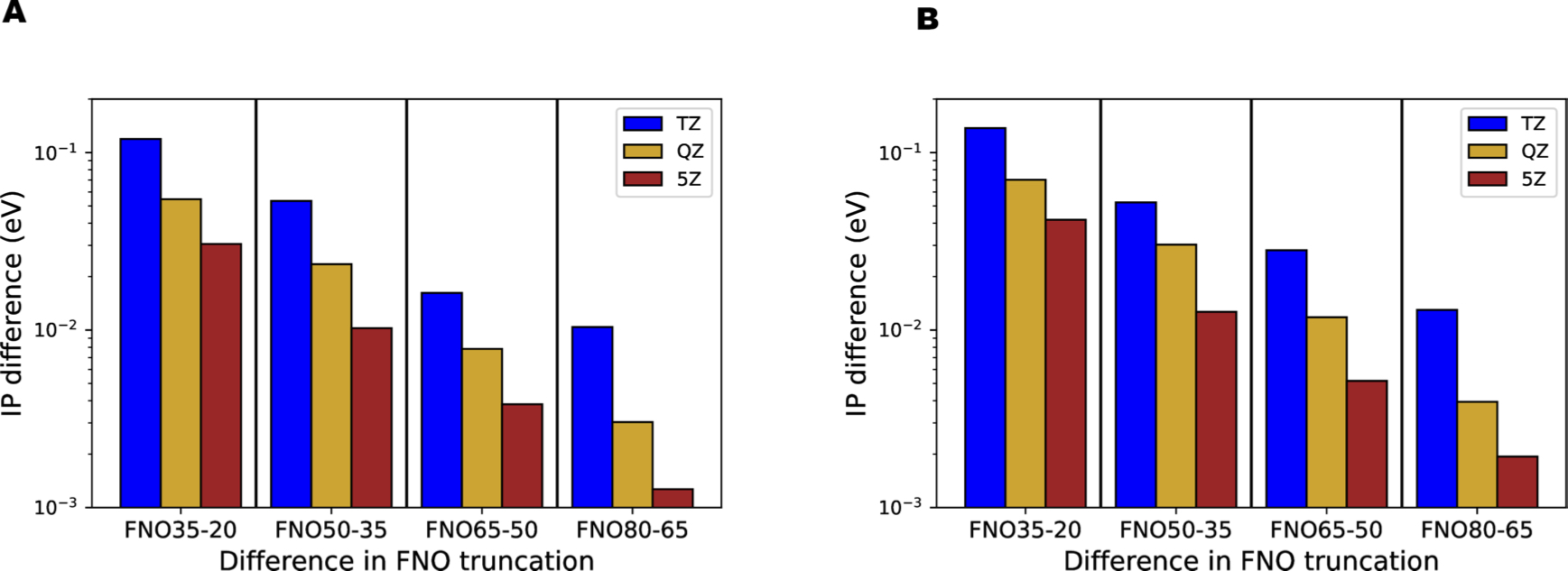}
\caption{Illustrating the impact successive FNO approximations have on the IP for A) ethlyene and B) 1,3 butadiene. The y-axis quantifies the difference in IP when using FNOs that keep 35\% of the virtual orbitals (VOs) versus FNOs that keep 20\% of the VOs, for example (e.g. FNO35-20), in the TZ, QZ, and 5Z basis sets.     }
\label{fig:FNObar}
\end{figure}

 We have thusfar shown that as we increase the number of VOs in the FNO approximation, the FNO-approximated IP value converges toward the unapproximated value. Ergo, we should expect that the difference between successive FNO approximation should asymptotically converge toward zero in the limit of FNOs that keep 100\% of the VOs. What role does oligomer size have in the FNO approximation error? To lend some perspective on this, Figure \ref{fig:FNObar} compares the error in IP between successive FNO approximations for ethelyene and 1,3-butadiene using the TZ, QZ, and 5Z basis sets. The overall error between successive FNO-based IP approximations slightly increases in the case of 1,3 butadiene, and is perhaps most notable when keeping 80\% versus 65\% of the VOs in the 5Z basis set. When considering the expediency of FNO-based calculations, this is a managable trade-off.

\section{Conclusions and future work}
% Discuss discontinuity in polyacene, area for future work using triples and FNOs
% Re-iterate the 'best', final polyacene and polyacetylene extrapolated band gaps
% Re-iterate the associated error of FNO
In spite of the FNOs being constructed with respect to the ground state reference and therefore not entirely rigorous for the determination of EAs, we have shown that errors with respect to the full EA-EOM-CCSD results are manageable when larger virtual spaces are kept. Alternatively, the FNO-EOM-IP predictions rapidly converge to the full EOM-IP results. Further analysis of the basis set convergence toward the CBS limit prove that EOM-IPs can be extracted using FNOs without compromising accuracy. 
Using FNOs that keep 35-40\% of the virtual space, predicted band gaps are within 0.1 eV of the full calculation.

This work showcases the utility in using FNOs for IP and EA calculations. As an example, their use has directly facilitated exploration of the band gap discontinuity phenomena present in polyacene that would have otherwise been computationally impractical to explore using more traditional wavefunction-based approaches. In this regard, we uncovered the existence of a low-lying, excited state surface of polyacene as the oligomer length increases beyond nonacene. We find that this higher-energy excited surface is in close proximity to the ``correct" ground state energy surface, and the band gap discontinuity previously reported is an artifact of mean-field theory's inability to properly differentiate between these electronic states in larger polyacene oligomers. Additional analysis of the electronic states originating from the several RHF$\rightarrow$UHF wavefunction instabilities could be facilitated by use of  ``template" QRHF orbitals.\cite{rittby1988open,pavlicek2024comparison}

Clearly, the use of FNOs are a pragmatic way of reducing EOM-CC calculation cost without significant sacrifice in accuracy. A point of future inquiry might involve adapting the meaning of FNOs to be more suitable for EAs and EEs. A possible way to do this involves picking one SR excited state and using it to generate a new set of FNOs which should then be amenable to EAs/EEs. Other ways undoubtedly exist to generate FNOs tailored for EE/EA states, and are worthy of further exploration.

 Any EOM-CC calculation has two steps. One is generation of $\overline{H}$ that comes solely from the ground reference state CC result. The second is the solution for the target state. The ground state MBPT2 is common to all states. But it would be possible to further tailor the FNO's to the target states if one wants some kind of ultimate accuracy for EA, IP, EE, DIP, DEA, etc states.
In each case, one has an MBPT2 approximation whose specific FNO's would likely allow for more basis set truncation, to provide close to basis set limits for the target state. With respect to band gap calculations, one has to ensure that both the target IP and EA are treated equivalently. One way to achieve this would be to pick a point equidistant between an IP and EA, analogous to a Slater Transition State, and then build MBPT2 from a CC calculation for that mid-point approximation. As in Slater's TS, this can be accomplished by averaging the density dependent J-K terms for the two limits, that then will generate a single determinant intermediate approximation for a subsequent correlated treatment. If one wants to treat several gaps equivalently it would be recommended to use the averaging procedure  in MBPT for a $\overline{V}(n-1)$ potential along with a $\overline{V}(n+1)$  potential.\cite{bartlett1974correlation} Now, presumably, the FNO's obtained at MBPT2 will be ideally suited to the direct calculation of the bang-gap.

%In summary, we find potential for using high-level wave function theory calculations in conjunction with FNOs on oligomeric, $\pi$-conjucated systems at or near the thermodynamic limit. 
\section*{Acknowledgements}
This work was supported by the Air Force Office of Scientific Research under AFOSR Award No. FA9550-23-1-0118. Z.W thanks the National Science Foundation and the Molecular Sciences Software Institute for financial support under Grant No. CHE-2136142. Z.W also acknowledges support from the U.S. Department of Energy, Office of Science, Office of Workforce Development for Teachers and Scientists, Office of Science Graduate Student Research (SCGSR) program. The SCGSR program is administered by the Oak Ridge Institute for Science and Education (ORISE) for the DOE. ORISE is managed by ORAU under contract number DE-SC0014664. 

\section*{References}

\bibliography{to_arxiv/main}

\end{document}